\documentclass[nofootinbib,aps,reprint]{revtex4-1}
\usepackage{graphicx}
\usepackage{dcolumn}
\usepackage{bm}
\usepackage{slashed}
\usepackage{amsmath}
\usepackage{caption}
\usepackage{subcaption}
\usepackage{float}
\usepackage{color}

\newcommand{\bal}{\nopagebreak[3]\begin{aligned}}
\newcommand{\eal}{\end{aligned}}

\DeclareMathAlphabet{\mathfs}{U}{rsfs}{m}{n}                     %

\newcommand{\n}{\nonumber}
\newcommand{\be}{\nopagebreak[3]\begin{equation}}
\newcommand{\ee}{\end{equation}}
\newcommand{\bee}{\nopagebreak[3]\begin{equation*}}
\newcommand{\eee}{\end{equation*}}
\newcommand{\ba}{\nopagebreak[3]\begin{eqnarray}}
\newcommand{\ea}{\end{eqnarray}}
\newcommand{\baa}{\nopagebreak[3]\begin{eqnarray*}}
\newcommand{\eaa}{\end{eqnarray*}}
\newcommand{\bseq}{\nopagebreak[3]\begin{subequations}}
\newcommand{\eseq}{\end{subequations}\noindent}

\begin{document}
\title{The First Law for the Kerr-NUT Spacetime}

\author{Ernesto Frodden}
 \email{efrodden@gmail.com}
\affiliation{Departamento de F\'isica, Universidad de Santiago de Chile, Avenida Ecuador 3493, Santiago, Chile.}

\author{Diego Hidalgo}
 \email{dihidalgot@gmail.com}
\affiliation{
 Centro de Estudios Cient\'ificos (CECs), Av. Arturo Prat 514, Valdivia, Chile\\
Departamento de F\'isica, Universidad de Concepci\'on, Casilla 160-C, Concepci\'on, Chile, } 
\affiliation{Instituto de Ciencias F\'isicas y Matem\'aticas, Universidad Austral de Chile, Casilla 567, Valdivia, Chile.
}

\date{\today}

\begin{abstract}

We show that the first law for the Kerr-NUT is straightforwardly established with the surface charge method. The entropy is explicitly found as a charge and its value is not proportional to the horizon area. We conclude that there are unavoidable contributions from the Misner strings to the charges, still the mass and angular momentum gets standard values. However, there are no independent charges associated with the Misner strings.  
\end{abstract}


\maketitle

\section{Introduction}\label{intro}

Taub-NUT (TN) spacetime is a vacuum solution of General Relativity \cite{Taub,NUT}. For a long time, the physical interpretation of TN spacetime has been elusive. This spacetime contains unusual properties that make it attractive for study and constitute a typical counter-example for weakly established statements in the field (think about {\it the area law} in black hole physics or standardized asymptotic behavior for spacetimes).~The characterizing feature of TN spacetime is that the additional parameter, the NUT parameter $n$, causes string-like defects \cite{Misner:1963fr}, nowadays called Misner strings. These strings are interpreted as the gravitational analogue of the Dirac strings appearing for magnetic monopoles in the electromagnetic context. Thus, the NUT property is further interpreted as a gravitomagnetic charge, {\it i.e.} as a charge dual to the mass in the same vein of electromagnetic duality.   

There are at least two remarkable interpretations of the TN solution. The first one, presented by Misner \cite{Misner:1963fr}, argued that the string-like defect could be made invisible if the time coordinate is made periodic. The second interpretation, employed by Bonnor \cite{bonnor1969}, keeps the strings as a certain ``massless source of angular momentum.'' 

Most of the studies on TN spacetimes is done in the Euclidean regime \cite{Chamblin:1998pz,Ghezelbash:2008zz,Emparan:1999pm,Mann:1999pc,Mann:1999bt,Garfinkle:2000ms,Johnson:2014xza,Johnson:2014pwa,Clarkson:2002uj,Astefanesei:2004kn,Kalamakis:2020aaj,Ciambelli:2020qny,Arratia:2020hoy,Hawking:1998jf,Hawking:1998ct}
 because gravitational instantons \cite{Hawking:1976jb} might play a role in the path integral formulation of gravity. However, the Lorentzian version of TN spacetime was critically revisited \cite{Clement:2015cxa,Clement:2015aka}. The authors considered the geodesics motion in the presence of Misner strings. The main result is that TN spacetime turns out to be geodesically complete: {\it i.e} Free-falling observers {\it do not see} strings as they cross them. Guided by this last analysis, renewed interest in TN spacetime emerged in the community as there is a window for the TN solutions to have physical relevance \cite{Kubiznak:2019yiu,Gonzalez:2017sfq,Cano:2021qzp,BallonBordo:2019vrn,Bordo:2019slw,Durka:2019ajz,Bordo:2019rhu,Bordo:2019tyh,Zhang:2021pvx}. 

In this work, we pursue a consistent formulation of thermodynamics for the Kerr-Nut spacetime, also known as {\it rotating} TN spacetime in the Lorentzian regime. This solution is also referred to as the Kerr NUT solution because it is a generalization of the Kerr black hole solution including the NUT parameter. The final hope is to settle down a thermodynamics interpretation for the TN spacetime to increase the knowledge about this maybe-physical solution and may help to have a new arena where quantum gravity ideas might be explored. As we will see, the spacetime entropy for this solution is different to the usual black holes.

Recently, \cite{Kubiznak:2019yiu}, it has been claimed that the first law for Lorentzian rotating TN spacetimes, Kerr-NUT, can be formulated assuming: 1) An entropy given by the one-four of the area of the horizon ({\it area law}), and 2) the existence of independently NUT parameters that control the effect of Misner strings at the first law.

In this work, we review how the Misner strings enter in the formulation of the first law. Using the surface charge method \cite{Frodden:2019ylc}. The TN charges: Mass and angular momentum, are derived with a suitable spacetime integration. With these conserved quantities, the variation of the entropy is naturally obtained as a charge too. This procedure, plus integrability considerations, lead us to restrict the parameter space and find an entropy that differs from the {\it area law}, in accordance with \cite{Hawking:1998jf}.

Our paper is organized as follows: In Section~\ref{thesolutiono}, we briefly review basic facts about the Kerr-NUT. The surface charge method improved  to deal with string-like defects is presented in Section~\ref{themethod}. Section~\ref{thethermo} is devoted to constructing the first law of thermodynamic for TN solution and find the appropriate entropy that fits on it.
As a consistency check of the results, in Subsection~\ref{ensemble}, we show that all exposed quantities and the first law can be framed in the grand canonical ensemble. Summary and final discussion are presented in Section~\ref{theconclu}. For completeness we include an appendix exhibiting intermediate steps of the computations.

\section{Kerr-NUT Solution}
\label{thesolutiono}
The Kerr-NUT, or rotating TN spacetime, generalizes the Kerr metric, with an extra NUT parameter $n$ as
\ba\label{thesolution}
\n && ds^ 2 =  -\frac{\Delta}{\Sigma} \left( dt +(2n\cos \theta +2Cn -a \sin^2\theta)d\phi \right)^2 + \frac{\Sigma}{\Delta} dr^2\\
&& +\,\frac{\sin^2\theta}{\Sigma} \left( a dt - (r^2+a^2+n^2-2anC)d\phi \right)^2 +\Sigma \, d\theta^2 \,,
\ea
with 
\ba
\Delta & =&  r^2+a^2 - 2m r -n^2  \,,\\
\Sigma & = &  r^2 +(n+a\cos \theta)^2\,,
\ea
where $m$ is the integration constant associated with the mass, $n$ the NUT parameter, and $a$ the rotation parameter. The solution with $C=0$ is a particular Demianski-Newman vacuum spacetime \cite{metriccite}. The parameter $C=\pm 1, 0$ was first introduced in \cite{Manko:2006bf}. The solution presents Misner strings as $(\nabla t)^2$ becomes ill-defined along the axis \cite{Misner:1963fr}. The discrete parameter $C$ specifies different configurations for the strings \cite{Manko:2006bf}:  For $C=+1$ one string is at the north pole axis while the south pole axis is regular, for $C=-1$ is the opposite situation, and for $C=0$ there are two strings symmetrically located. 

The solution has a Killing horizon \cite{BallonBordo:2019vrn,podolskybook} located at the largest root of $\Delta=0$ and it is generated by the Killing vector $\xi=\partial_t+\Omega\, \partial_\phi$, with $\Omega$ the angular velocity of a ZAMO (zero angular momentum observer) at the horizon as seen by a static observer at infinity.
\section{Surface charges for spacetimes with strings}\label{themethod}
To characterize TN spacetime we can compute charges associated with isometries. General Relativity is a gauge theory and a naive use of the First Noether Theorem produces misleading charges. An improved method to extract charges is required. We use the standard surface charge method \cite{Lee:1990nz,Iyer:1994ys,Wald:1999wa}; for isometries it is equivalent to the symplectic method explained in \cite{Barnich:2001jy} (both methods were revisited in \cite{Frodden:2019ylc}).
 
An exact symmetry (here equivalent to a Killing symmetry) that is generated by a Killing vector field $\xi$ produces an on-shell conservation law equation: $d \boldsymbol{k}_\xi\approx 0$. The integration of this equation over a spacetime volume $\Sigma$, with two disconnected boundaries $\partial \Sigma=S \cup S^{\prime}$, defines a {\it conserved} surface charge variation as
\begin{equation}
 \slashed{\delta} Q_{\xi}=   \oint_{S} \boldsymbol{k}_\xi \,, 
\label{scintegral}
\end{equation}
the conservation is understood as the fact that for any other closed two-surface $S^{\prime}$, obtained from a continuous deformation of $S$; the previous expression produces the same $\slashed{\delta} Q_{\xi}$. In particular, this is true for deformations in time and radial directions. 
 In' \eqref{scintegral} we use bold notation to indicate spacetime differential forms, namely the surface charge density is the two-form $\boldsymbol{k}_\xi=  \frac{1}{4}k^{\mu \nu}_{\xi} \varepsilon_{\mu \nu\rho\sigma} dx^\rho \wedge dx^\sigma$. See  \cite{Frodden:2019ylc}  for more details and the explicit formula of $k^{\mu \nu}_{\xi} $ in terms of a general metric or tetrad variables.

Here we stress two aspects of \eqref{scintegral}. First, the slashed delta symbol in $ \slashed{\delta} Q_{\xi}$ means we compute a one-form in phase space which is not necessarily exact. Exactness in phase space, namely $ \delta (\slashed \delta Q_\xi)=\delta^2Q_\xi=0$, is an extra condition required for the existence of the charge $Q_\xi$, in the usual parlance this is the {\it integrability condition}. A second aspect is that integration is over any closed smooth surface $S$ where the conservation law holds (not necessarily at {\it infinity}). As we check in a moment, Misner strings violate this requirement, and the integration region needs to be corrected accordingly.     

As in Kerr spacetime, for the Kerr-NUT, there are also two exact symmetries present: Time translations generated by $\partial_t$ and axial rotations generated by $\partial_\phi$. A direct computation of \eqref{scintegral} fails as it produces charges dependent on the surface $S$ ({\it e.g.} dependent on the radius of a given sphere): Conservation is broken! This is a problem the reader may be tempted to solve by taking a sphere at {\it infinity}. But, as explained before, conservation here means precisely that one can compute charges not necessarily at the asymptotic region. The solution to the problem comes by excluding the region with the Misner strings. More precisely, we integrate the conservation law equation in a volume without the strings. As a consequence, the result does not decouple into the usual two integrals, one over each closed surface $S^\prime$ and $S$ (which is what makes of \eqref{scintegral} a seed to compute a charge). Instead, extra contributions to \eqref{scintegral} emerge.  

\begin{figure}[h!]
    \centering
    \includegraphics[width=7cm]{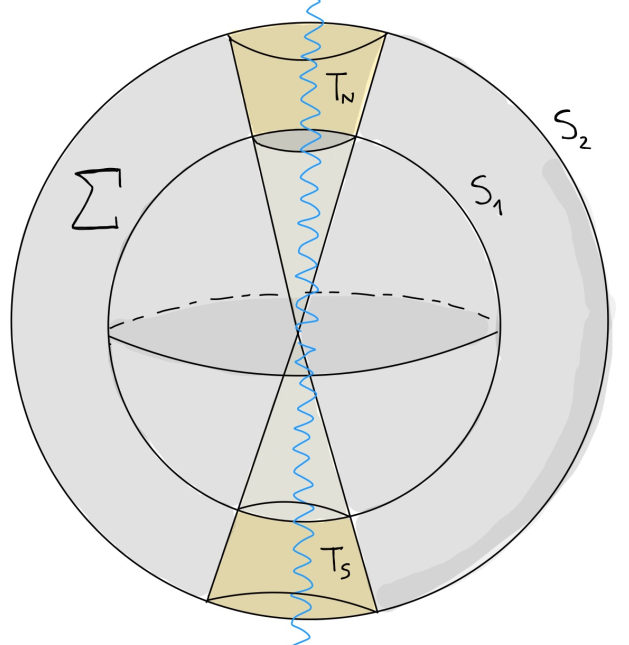}
    \caption{Integration region: A three-volume, $\Sigma$, delimited by two spheres and two cones. The blue lines stand for the Misner strings symmetrically positioned (case $C=0$).}
    \label{figureintegration}
\end{figure}

Let us consider the surface charge density $\boldsymbol{k}_{\xi}$, with $\xi$ a Killing vector field, either $\partial_t$ or  $\partial_\phi$. We integrate the three-form $d\boldsymbol{k}_{\xi}$ on a volume $\Sigma_\epsilon$ limited by two spherical shells, say $S_1$ and $S_2$, with radius $r_1<r_2$, and two tubes $T_S$ and $T_N$ (actually cones) excluding the strings from the volume, see Figure \ref{figureintegration}. The south cone $T_S$ has a small opening angle, $\theta =\pi -\epsilon$, towards the south pole, and the north cone $T_N$ has a small opening angle, $\theta =0+\epsilon$, towards the north pole. We integrate $d\boldsymbol{k}_{\xi}\approx 0$ over $\Sigma_\epsilon$, thus
\be
\label{conservationlawTaub}
0= \int_{\Sigma_\epsilon}\!\! d \boldsymbol{k}_{\xi} =\int_{\partial \Sigma_\epsilon}\!\!  \boldsymbol{k}_{\xi}=\int_{S_1}\!\! \boldsymbol{k}_{\xi} -\int_{S_2}\!\! \boldsymbol{k}_{\xi}   + \int_{T_S}\!\!\! \boldsymbol{k}_{\xi}  +\int_{T_N}\!\!\! \boldsymbol{k}_{\xi}\,.
 \ee
When there are no strings, the third and fourth term vanishes in the shrinking limit $\epsilon\!\to\! 0$ and we recover the usual expression for the charge \eqref{scintegral}. However, for TN spacetime, strings contribute in the limit of a vanishing $\epsilon$.  

In general, the radial integration over each tube decouples into two terms associated with the circles found as the intersection of the tube with the shells 
\bseq\label{tubos}
\ba
\int_{T_S} \boldsymbol{k}_\xi & =&  K_{\xi}\big|_{r=r_1; \,\theta=\pi-\epsilon} - K_{\xi}\big|_{r=r_2; \,\theta=\pi-\epsilon}\,,\\
\int_{T_N} \boldsymbol{k}_\xi & =&  K_{\xi}\big|_{r=r_1; \, \theta=\epsilon} - K_{\xi}\big|_{r=r_2; \,\theta=\epsilon}\,.
\ea
\eseq
Then, we take the limit $\epsilon\to 0$ and reorder \eqref{conservationlawTaub} with terms associated with each shell to get an expression for the surface charge that is defined exclusively at a sphere $S=S_1$. This quantity is truly conserved, {\it i.e.} its value does not depend on the specific surface $S$ 
\be\label{chargetaubnutgeneral}
\slashed{\delta}Q_\xi = \oint_{S} \boldsymbol{k}_{\xi} + K_{\xi}\big|_{south} +   K_{\xi}\big|_{north}\,,
\ee
where we refer to $\theta=\pi$ as south, and $\theta=0$ as north. Although we formally use a specific parametrization to refer to geometrical objects, the result does not depend on them. Before shrinking the latter around the strings, we can deform either the shells or the tubes and still have the same resulting surface charge. In particular, \eqref{chargetaubnutgeneral} does not depend on the specific radius we compute it. 

Therefore, we have shown how the $K$ contributions solve the problem of a naive computation with \eqref{scintegral} when string-like defects cross the surface $S$. For spacetimes without strings the $K$ terms automatically vanish. Still, we note that we can introduce artificial string-like defects by changing coordinates even for regular spacetimes. This can break down the conservation of \eqref{scintegral} unless terms similar to the $K$ are taken into account. An example of this is the defect appearing in the axis for some versions of the Kerr-(a)dS metric solution which can be avoided by a $\phi$ escalation.  

We emphasize that to get \eqref{chargetaubnutgeneral}, we rely on the correct integration of the local conservation law. This requires identifying the tubes to exclude the defects: The regions where  $d \boldsymbol{k}_\xi\approx 0$ does not hold.

\subsection{Mass and Angular Momentum }

The two Killing symmetries produce two independent varied surface charges. We check that our corrected formula delivers the expected result.

The solution space is spanned by three parameters $m$, $a$, and $n$. For time translations exact symmetry, we expect the mass/energy, thus we rename $\delta M\equiv \slashed\delta Q_{\partial_t}$, and \eqref{chargetaubnutgeneral} produces (for details check the appendix)
\be
\delta M \, = \,  \delta m \,,
\ee  
which is of course integrable as $\delta^2M=0$. We set the integration constant equal to zero to obtain the finite charge $M =  m$, which means that for Kerr-NUT spacetime the parameter $m$ is still identified with the total mass/energy of spacetime.

For the axial symmetry generator we expect the angular momentum, thus we rename $\delta J\equiv \slashed\delta Q_{-\partial_\phi}$, and we get
\be\label{deltaJ}
\delta J \, = \, (a-3Cn )\delta m+m \delta a-3Cm\delta n \,.
\ee
which again is an integrable expression, $\delta^2J=0$, and produces the angular momentum $J = (a-3Cn )m$. Note that, as expected, the TN spacetime with $a=0$ still carries angular momentum but only when strings are not symmetrically distributed along the rotation axis. 

Let us conclude here with two remarks. First, the charges do not present divergences ({\it e.g.} $r$-divergences when evaluated in asymptotic regions). While each integral in \eqref{chargetaubnutgeneral} is $r$-dependent, therefore susceptible to divergences, it is only the sum of the three that produces $r$-independent charges. Then, the $K$-es are essentials and the Misner strings contribute to the charges though them. 
The second remark is that no asymptotic region is required. We do not demand a particular location for the sphere $S$. In that sense, the charges computed above are quasi-local.  

\section{Entropy and The First Law}\label{thethermo}
In analogy with black holes thermodynamics, we now treat the Kerr-NUT spacetime as a thermodynamic system. We constructively present the first law to determine the entropy of this spacetime following the black hole prescription, {\it i.e.} we introduce a temperature and an angular momentum related to the surface $r=r_+$, the larger root of $\Delta=0$. For later convenience, we focus on the case with $C=0$.

Let us introduce a temperature (proportional to the surface gravity at $r_+$ redshifted to infinity)
\be\label{temp}
T= \frac{\sqrt{m^2-a^2+n^2}}{4 \pi  \left(m \left(m+\sqrt{m^2-a^2+n^2}\right)+n^2\right)}.\,
\ee
The angular velocity, of the surface $r=r_+$ as measured from infinity, is
 \be\label{omega}
 \Omega=\left.\frac{g_{t\phi}}{g_{\phi \phi}}\right|_{r_+} = -\frac{a}{2 \left(m \left(m+\sqrt{m^2-a^2+n^2}\right)+n^2\right)}\, .
 \ee 
The expectation is that in the first law the temperature multiplies the entropy variation, we call it $\slashed{\delta} S$, and the angular velocity multiplies the angular momentum variation (\ref{deltaJ}). Then, we can isolate $\slashed{\delta} S$ from the first law to get  
\begin{multline}\label{Snonint}
 \slashed{\delta} S  =   \frac{1}{T} \left( \delta M- \Omega \delta J\right)=\frac{2\pi }{\sqrt{m^2-a^2+n^2}}\times\\
 \left(  \big(2 m (m+\sqrt{m^2-a^2+n^2})- a^2+2n^2\big)\delta m- m a  \delta a \right)\,.
\end{multline}

Note that for $n=0$ the previous expression is integrable, and as expected, it is the variation of the usual one-fourth of the area black hole entropy of Kerr spacetime. Even for non-rotating TN spacetime, $n\neq 0$, the expression is not integrable anymore, $\delta^2S\neq 0$, which is a problem to have a well-defined entropy charge. 

Still the entropy variation is conserved in the sense that it can be directly computed from \eqref{chargetaubnutgeneral} with the combination of Killing vectors $\xi=\frac{1}{T}\left(\partial_t+\Omega\, \partial_\phi\right)$. That means it is a proper varied surface charge as it does not depend on the surface one chooses to compute it. Therefore, this expression should be a good seed to compute an entropy as a Noether charge as we already did with the mass or the angular momentum (we keep the {\it Noether charge} name, understanding that surface charge method to compute charges generalizes the Noether procedure \cite{Frodden:2019ylc}).

In the literature, it has been explored the possibility of adding extra terms to the first law associated with the Misner strings. At this level, this strategy is equivalent to the splitting of \eqref{Snonint} in different terms and providing interpretation to each one. Here we choose another path. We use the surface charge method as a guide and keep the possibility of having an entropy that is indeed conserved: A Noether charge. Therefore the expression in \eqref{Snonint} must not be split; otherwise, we can not write the entropy variation as \eqref{chargetaubnutgeneral} with an arbitrary closed surface. 

{\it The alternative path}: We have checked the suggestive alternative of splitting \eqref{Snonint} into surface and tubes terms associated to the Misner strings as in \eqref{chargetaubnutgeneral}.~As explained before, separately surfaces and tubes terms are not conserved quantities and neither they are integrable.~One may be tempted to fix the closed surface at $r=r_+$. In this case, the surface integral in \eqref{chargetaubnutgeneral} indeed becomes the variation of one-fourth of the area. However, we are not computing a conserved charge if we have to fix the surface where it is defined. Furthermore, insisting in this approach, the terms associated with the tubes are not integrable, and another {\it temperatures} should be introduced to achieve their integrability. This path would be similar to the one presented in \cite{Bordo:2019rhu} (still, we have a different angular momentum stemming from the other methods used to compute charges).  

A way to have, simultaneously, the entropy as conserved and integrable is to reduce the parameter space. Let us assume that $m$, $a$, and $n$ are dependent quantities and solve this dependence such that the integrability condition is satisfied. Consider $m=m(n,a)$. Because of the way these parameters enter in the metric components, it is easy to see that all the tree parameters scale with the same factor under scaling of the metric. Then, the allowed relation among the parameters is simpler: $m=n\, h(a/n)$, with $h(a/n)$ a function to be determined by the integrability condition $\delta (\slashed \delta S)=0$. This produces a differential equation solved simply by $h(a/n)=\alpha$, with $\alpha$ an integration constant. Thus
\be
m=\alpha\,  n \,,
\ee
 ensures integrability of the entropy, and it reduces the parameter space of our solution from three to two parameters. Replacing this in the equation for the varied charge \eqref{Snonint}, we can integrate it to obtain an entropy
\be\label{entropy}
S = 2 \pi  \alpha\,  n^2 \left(\alpha  + \sqrt{1+\alpha ^2 -\frac{a^2}{n^2}}\right)\,,
\ee
where we have set  to zero the integration constant. Our method does not fix $\alpha$ and in what follows we keep it free.\footnote{We can fix $\alpha$ if we require the entropy to be compatible with the non-rotating TN entropy obtained through other methods. For instance, Hawking and Hunter in \cite{Hawking:1998jf} found for the Euclidean TN-bolt counterpart $S_{HH}=\pi n^2$, this fixes $\alpha= \frac{1}{2\sqrt{2}}$.} 
This entropy fulfills a standard first law
\be
\delta M = T \delta S + \Omega \delta J\,,
\ee  
where $M=  \alpha\, n$ and $J=  \alpha\, a\, n$. In contrast with the usual black hole solutions, the entropy is not one-fourth of the area, first noted in \cite{Hawking:1998jf} for non-rotating TN and using Euclidean methods. The first law presented is minimal because it does not require additional terms not justified in terms of charges. It may be seen as non-standard as we had reduced the parameter space to achieve integrability. But as explained before, this is required to have $M$, $J$, and $S$ as well-defined charges to compose the first law. Therefore, we conclude that there is a consistent thermodynamic description only for a subset of parameter space of the Kerr-NUT spacetime. If TN spacetimes will ever have a physical meaning this is a prediction of our approach.
\subsection{Grand Canonical Ensemble}\label{ensemble}
Our result is compatible with the Euclidean action to describe the thermodynamics. To show this, we use the grand canonical ensemble function, $G$, already computed in the Appendix of \cite{Bordo:2019rhu,Emparan:1999pm} as the flat limit of a regularized Euclidean (a)dS action divided by $\beta=1/T$. The result is simply $G=m/2$. We express it as a function of the intensive thermodynamic variables at play to get
\be
G(T,\Omega)=\frac{1}{8\pi T \left(  1 + \alpha^\prime \sqrt{  1+\displaystyle \frac{\Omega^2}{4\pi^2T^2}}  \right)}\,.
\ee
With $\alpha^\prime=\sqrt{1+1/\alpha^2}$. The expression is consequence of a simple algebraic manipulation of $T(a,n)$, $\Omega(a,n)$, and $m=\alpha\, n$.  Note its simplicity. Now, the first law for the Kerr-NUT spacetimes can be reproduced in the standard way. As expected,
the entropy \eqref{entropy}, the angular momentum, and the mass/energy are: $S=-\frac{\partial G}{\partial T}$, $J=\frac{\partial G}{\partial \Omega}$ and $M=G- J\Omega+ST $, respectively. 

\section{Conclusions and Outlook}\label{theconclu} 

In this note, we reported a procedure to use the covariant surface charge method for spacetimes with string-like defects and immediately applied it to the Kerr-NUT spacetime. The critical step was to note that the equation, $d{\boldsymbol k}_\xi\approx 0$, is not well behaved over the Misner strings; therefore, the three-volume where it is integrated can not contain them. The procedure is general and can be easily applied in particular for accelerating black holes.

For the Kerr-NUT spacetime, besides the explicit computation of the natural charges, mass and angular momentum, the thermodynamic picture was completed by writing down the first law and computing the entropy as an integrated charge. We achieve the integrability of the entropy without introducing additional terms in the first law. As a general rule, the introduction of new terms in the first law should be done with caution, we try to avoid it. Instead, we impose a reduction in the parameter space of the solution. Thus, the principle that charges appearing in the first law ($M$, $J$, and $S$) are Noether charges,  all at the same footing, is respected. The entropy found is not proportional to the area, which is something already expected for TN spacetimes.

Let us finish with a comment on our attempt to extend the results to the (a)dS Kerr-NUT family. The strategy to compute mass and angular momentum charges is equally successful there, however we could not solve the differential equation for the integrability of the entropy. Then, we leave for a future project to extend the thermodynamic picture for that spacetime family or in full generality for the Plebanski-Demianski family \cite{Griffiths:2005qp}.

\begin{acknowledgments}
The authors would like thank to Remigiusz Durka for a valuable discussion on the topic.
The Centro de Estudios Cient\'{\i}ficos (CECs) is funded by the Chilean
Government through the Centers of Excellence Base Financing Program of ANID. 
\vspace{0.1cm}

\end{acknowledgments}

\appendix
\section{Charge Computations}
The surface charge for GR is given by  \cite{Lee:1990nz,Iyer:1994ys,Wald:1999wa,Barnich:2001jy,Frodden:2019ylc}
\ba
\n k_{\xi}^{\mu\nu}& =&\sqrt{-g}\kappa\left(\xi^{[\nu}\nabla_\sigma \delta g^{\mu]\sigma}-\xi^{[\nu}\nabla^{\mu]} \delta g+\xi_\sigma\nabla^{[\mu} \delta g^{\nu]\sigma} \right. \\
&& ~~~~~~~~~\left.-\frac{1}{2}\delta g\nabla^{[\nu}\xi^{\mu]}+\delta g^{\sigma[\nu}\nabla_\sigma\xi^{\mu]}\right)\,,
\ea
where $\kappa = c^4/(8\pi G)$, $c$ the speed of light, $G$ the Newton's constant, $g$ stands for the determinant of the metric $g_{\mu \nu}$, and $\delta g\equiv g_{\mu\nu}\delta g^{\mu\nu}$.\footnote{We work in units where $c=G=1$.} Notice that the last three terms of the equation corresponds to the variation of the Komar's integrand commonly used to compute charges. 

The integral of $k_{\xi}^{\mu\nu}$ over a closed surface splits in four regions: Two concentric spheres with holes  and two cones (see Fig.~1). Over a sphere at constant $r$ the required component is $k^{tr}_{\xi}$ while for the cones it is $k^{\theta t}_{\xi}$. The explicit expressions for them in the Kerr-Taub-NUT spacetime are quite large, thus, for the sake of space with do not present them here (still it is a straightforward computation using a software). To give a taste we exhibit just the coefficient accompanying the $\delta m$ term of $k_{\partial_{t}}^{t r} \equiv k^{tr}_{(t)}$ associated to a Killing vector field $\xi = \partial_t$. That is, we decompose 
\be
k^{tr}_{(t)}=\left[k^{tr}_{(t)}\right]_{\delta m}\delta m+\left[k^{tr}_{(t)}\right]_{\delta a}\delta a+\left[k^{tr}_{(t)}\right]_{\delta n}\delta n\,,\n
\ee
and the first term is explicitly 
\begin{multline}
\left[k^{tr}_{(t)}\right]_{\delta m}=\\
\frac{\sin\theta}{8\pi}\left(1+\frac{\left(r^2+a^2-2 a nC+n^2\right) (r^2-a^2 \cos^2 \theta )}{\left(r^2+ a \cos \theta  (a \cos \theta +2 n)+n^2\right)^2}\right) \,. \n
\end{multline}
Note that it is a function of the coordinates. Here we can already identify the first term that will contribute to {\it half} of the final charge $m$. We can also check why, for this particular charge, the asymptotic analysis works as in the limit $r\to \infty$ the second term in the parenthesis reduces to 1. 

Before continuing with the calculation, let us define some functions that will allow us to  compactly write our next expressions 
\ba
h_{\pm}&=& (a \pm n)^2 + r^2 \,, \\
f_{\delta a_{\pm}}&=& n (-m (a \pm n)^2 \pm 2 n (a \pm n) r + m r^2)\,, \\
\n f_{\delta n_{\pm}}&=& m (a \pm n)^2 (2 a \pm n) \mp n (a \pm n) (3 a \pm n) r \\
~~ ~~&& +\,  m (2 a \pm 3 n) r^2 \mp 3 n r^3\,.
\ea

For the Killing vector field $\xi=\partial_t$, the terms from the cones contributing to the charges in \eqref{tubos} are
\ba
\int_{T_S} \boldsymbol{k}_{(t)} & =&  \int_{r_1}^{r_2} \hspace{-0.2cm} dr \int_0^{2\pi} \hspace{-0.2cm}d\phi \,   k^{\theta t}_{(t)} \bigg|_{\theta=\pi-\epsilon}  \hspace{-0.6cm} = K_{(t)}\bigg|_{\substack{r=r_1, \\\theta=\pi-\epsilon}} \hspace{-0.2cm}\hspace{-0.3cm}- K_{(t)}\bigg|_{\substack{r=r_2 ,\\ \theta=\pi-\epsilon}}\,,\\
\int_{T_N} \boldsymbol{k}_{(t)} & =&\int_{r_1}^{r_2} \hspace{-0.2cm}dr \int_0^{2\pi}\hspace{-0.2cm} d\phi  \, k^{\theta t}_{(t)}\bigg|_{\theta=\epsilon} \hspace{-0.2cm}=   K_{(t)}\bigg|_{\substack{r=r_1,\\ \theta=\epsilon}} \hspace{-0.1cm}\hspace{-0.3cm} - K_{(t)}\bigg|_{\substack{r=r_2 ,\\ \theta=\epsilon}}\,,
\ea
where $\epsilon$ is the small opening angle for the cones. We take $\epsilon\to 0$ after integrating on each surface. Note that we have included an explicit intermediate term, and note that for that term $\theta$ names a component of the tensorial density $k^{\mu\nu}_{(t)}$ but it is also a coordinate where the spacetime function $k^{\mu\nu}_{(t)}$ can be evaluated. We concentrate in the region with $r_2\to r$, then the two ending terms of the last equations are, explicitly, given by 
\ba
K_{(t)}\big|_{north}&=&\lim_{\epsilon\to 0}K_{(t)}\big|_{\substack{r ,\\ \theta=\epsilon}}\\
&=& \frac{C+1}{2 h_{+}^2} \left( n (a + n)h_{+}\delta m+f_{\delta a_{+}}\delta a  + f_{\delta n_{+}}
\delta n\right)\,,\n\\
 K_{(t)}\big|_{south}&=& \lim_{\epsilon\to 0}K_{(t)}\big|_{\substack{r,\\ \theta=\pi-\epsilon}}\\
&=& \frac{C-1}{2h_{-}^{2}} \left(n (a - n)h_{-}\delta m+f_{\delta a_{-}}\delta a  + f_{\delta n_{-}}
\delta n\right)\,.\n
\ea

For the sphere contribution we have
\ba
\oint_{S} \boldsymbol{k}_{(t)}&=&\lim_{\epsilon\to 0}\left[\int_{\epsilon}^{\pi-\epsilon} \hspace{-0.2cm}d\theta \int_0^{2\pi} d\phi\  k^{tr}_{(t)}\right]\\ 
&=&\left(1-\frac{(C-1) n (a-n)}{2h_-}-\frac{(C+1) n (a+n)}{2h_+}\right)\delta m\n\\
&&-  \left(\frac{(C-1) f_{\delta a_-}}{2h_-^2}+\frac{(C+1) f_{\delta a_{+}}}{2h_+^2}\right)\delta a\n\\
&&- \left(\frac{(C-1) f_{\delta n_-}}{2h_-^2}+\frac{(C+1) f_{\delta n_{+}}}{2h_+^2}\right) \delta n\,.\n
\ea
Finally, the variation of the charge reads (see Eq.~\eqref{chargetaubnutgeneral})
\be
\delta M\equiv \slashed{\delta}Q_{(t)} = \oint_{S} \boldsymbol{k}_{(t)} + K_{(t)}\big|_{south} +   K_{(t)}\big|_{north}=\delta m\,,
\ee
which happens to be trivially integrable. We see that most of the contributions to the charge nicely cancel such that the final result does not depend on coordinates (here the radius of the sphere). Note how complicated the $r-$dependency of each term may be. Still, the whole charge is independent as the method ensures conservation by virtue of $\partial_\mu k^{\mu\nu}_{(t)}\approx 0$.\\
The computation of the (varied) angular momentum $\delta J\equiv\slashed \delta Q_{-\partial_\phi}= \delta\left((a-3C n)m\right)$ associated with the Killing vector field $\xi=-\partial_\phi$ can be carried out following the same steps as the derivation of the mass.



\begin{thebibliography}{100}

\bibitem{Taub}
A. H.~Taub,~ {\it Empty space-times admitting a three parameter group of motions, Annals of Mathematics} {\bf 53} 472-490 (1951).
 
\bibitem{NUT}
E.~Newman, L.~Tamburino, and T.~Unti, {\it Empty-space generalization of the Schwarzschild metric, Journal of Mathematical Physics} {\bf 4} (1963), no. 7 915–923.

\bibitem{Misner:1963fr}
C.~W.~Misner,
``The Flatter regions of Newman, Unti and Tamburino's generalized Schwarzschild space,''
J. Math. Phys. \textbf{4}, 924-938 (1963)
doi:10.1063/1.1704019

\bibitem{bonnor1969}
Bonnor,~W.~B.~ 1969 {\it Proc. Camb. Phil. Soc. {\bf 66} 145}

\bibitem{Chamblin:1998pz}
A.~Chamblin, R.~Emparan, C.~V.~Johnson and R.~C.~Myers,
``Large N phases, gravitational instantons and the nuts and bolts of AdS holography,''
Phys. Rev. D \textbf{59}, 064010 (1999)
doi:10.1103/PhysRevD.59.064010
[arXiv:hep-th/9808177 [hep-th]].
Copy to ClipboardDownload
\bibitem{Ghezelbash:2008zz}
A.~M.~Ghezelbash, R.~B.~Mann and R.~D.~Sorkin,
Can. J. Phys. \textbf{86}, 579-582 (2008)
doi:10.1139/p07-184

\bibitem{Emparan:1999pm}
R.~Emparan, C.~V.~Johnson and R.~C.~Myers,
``Surface terms as counterterms in the AdS / CFT correspondence,''
Phys. Rev. D \textbf{60}, 104001 (1999)
doi:10.1103/PhysRevD.60.104001
[arXiv:hep-th/9903238 [hep-th]].

\bibitem{Mann:1999pc}
R.~B.~Mann,
``Misner string entropy,''
Phys. Rev. D \textbf{60}, 104047 (1999)
doi:10.1103/PhysRevD.60.104047
[arXiv:hep-th/9903229 [hep-th]].

\bibitem{Mann:1999bt}
R.~B.~Mann,
``Entropy of rotating Misner string space-times,''
Phys. Rev. D \textbf{61}, 084013 (2000)
doi:10.1103/PhysRevD.61.084013
[arXiv:hep-th/9904148 [hep-th]].

\bibitem{Garfinkle:2000ms}
D.~Garfinkle and R.~B.~Mann,
``Generalized entropy and Noether charge,''
Class. Quant. Grav. \textbf{17}, 3317-3324 (2000)
doi:10.1088/0264-9381/17/16/314
[arXiv:gr-qc/0004056 [gr-qc]].

\bibitem{Johnson:2014xza}
C.~V.~Johnson,
``Thermodynamic Volumes for AdS-Taub-NUT and AdS-Taub-Bolt,''
Class. Quant. Grav. \textbf{31}, no.23, 235003 (2014)
doi:10.1088/0264-9381/31/23/235003
[arXiv:1405.5941 [hep-th]].

\bibitem{Johnson:2014pwa}
C.~V.~Johnson,
``The Extended Thermodynamic Phase Structure of Taub-NUT and Taub-Bolt,''
Class. Quant. Grav. \textbf{31}, 225005 (2014)
doi:10.1088/0264-9381/31/22/225005
[arXiv:1406.4533 [hep-th]].

\bibitem{Clarkson:2002uj}
R.~Clarkson, L.~Fatibene and R.~B.~Mann,
``Thermodynamics of (d+1)-dimensional NUT charged AdS space-times,''
Nucl. Phys. B \textbf{652}, 348-382 (2003)
doi:10.1016/S0550-3213(02)01143-4
[arXiv:hep-th/0210280 [hep-th]].


\bibitem{Astefanesei:2004kn}
D.~Astefanesei, R.~B.~Mann and E.~Radu,
``Nut charged space-times and closed timelike curves on the boundary,''
JHEP \textbf{01}, 049 (2005)
doi:10.1088/1126-6708/2005/01/049
[arXiv:hep-th/0407110 [hep-th]].

\bibitem{Kalamakis:2020aaj}
G.~Kalamakis, R.~G.~Leigh and A.~C.~Petkou,
``Aspects of holography of Taub-NUT- AdS$_4$ spacetimes,''
Phys. Rev. D \textbf{103}, no.12, 126012 (2021)
doi:10.1103/PhysRevD.103.126012
[arXiv:2009.08022 [hep-th]].

\bibitem{Ciambelli:2020qny}
L.~Ciambelli, C.~Corral, J.~Figueroa, G.~Giribet and R.~Olea,
``Topological Terms and the Misner String Entropy,''
Phys. Rev. D \textbf{103}, no.2, 024052 (2021)
doi:10.1103/PhysRevD.103.024052
[arXiv:2011.11044 [hep-th]].

\bibitem{Arratia:2020hoy}
E.~Arratia, C.~Corral, J.~Figueroa and L.~Sanhueza,
``Hairy Taub-NUT/bolt-AdS solutions in Horndeski theory,''
Phys. Rev. D \textbf{103}, no.6, 064068 (2021)
doi:10.1103/PhysRevD.103.064068
[arXiv:2010.02460 [hep-th]].

\bibitem{Hawking:1998jf}
S.~W.~Hawking and C.~J.~Hunter,
``Gravitational entropy and global structure,''
Phys. Rev. D \textbf{59}, 044025 (1999)
doi:10.1103/PhysRevD.59.044025
[arXiv:hep-th/9808085 [hep-th]].

\bibitem{Hawking:1998ct}
S.~W.~Hawking, C.~J.~Hunter and D.~N.~Page,
``Nut charge, anti-de Sitter space and entropy,''
Phys. Rev. D \textbf{59}, 044033 (1999)
doi:10.1103/PhysRevD.59.044033
[arXiv:hep-th/9809035 [hep-th]].

\bibitem{Hawking:1976jb}
S.~W.~Hawking,
``Gravitational Instantons,''
Phys. Lett. A \textbf{60}, 81 (1977)
doi:10.1016/0375-9601(77)90386-3

\bibitem{Clement:2015cxa}
G.~Cl\'ement, D.~Gal'tsov and M.~Guenouche,
``Rehabilitating space-times with NUTs,''
Phys. Lett. B \textbf{750}, 591-594 (2015)
doi:10.1016/j.physletb.2015.09.074
[arXiv:1508.07622 [hep-th]].

\bibitem{Clement:2015aka}
G.~Cl\'ement, D.~Gal'tsov and M.~Guenouche,
``NUT wormholes,''
Phys. Rev. D \textbf{93}, no.2, 024048 (2016)
doi:10.1103/PhysRevD.93.024048
[arXiv:1509.07854 [hep-th]].



\bibitem{Kubiznak:2019yiu}
R.~A.~Hennigar, D.~Kubiz\v{n}\'ak and R.~B.~Mann,
``Thermodynamics of Lorentzian Taub-NUT spacetimes,''
Phys. Rev. D \textbf{100}, no.6, 064055 (2019)
doi:10.1103/PhysRevD.100.064055
\href{https://arxiv.org/pdf/1903.08668.pdf}{[arXiv:1903.08668 [hep-th]]}.
\bibitem{Cano:2021qzp}
P.~A.~Cano and D.~Pere\~niguez,
[arXiv:2101.10652 [hep-th]].
\bibitem{Gonzalez:2017sfq}
H.~Gonzalez, D.~Grumiller, W.~Merbis and R.~Wutte,
EPJ Web Conf. \textbf{168}, 01009 (2018)
doi:10.1051/epjconf/201816801009
[arXiv:1709.09667 [hep-th]].
\bibitem{BallonBordo:2019vrn}
A.~Ballon Bordo, F.~Gray, R.~A.~Hennigar and D.~Kubiz\v{n}\'ak,
``The First Law for Rotating NUTs,''
Phys. Lett. B \textbf{798}, 134972 (2019)
doi:10.1016/j.physletb.2019.134972
[arXiv:1905.06350 [hep-th]].

\bibitem{Bordo:2019slw}
A.~B.~Bordo, F.~Gray and D.~Kubiz\v{n}\'ak,
``Thermodynamics and Phase Transitions of NUTty Dyons,''
JHEP \textbf{07}, 119 (2019)
doi:10.1007/JHEP07(2019)119
[arXiv:1904.00030 [hep-th]].

\bibitem{Durka:2019ajz}
R.~Durka,
``The first law of black hole thermodynamics for Taub-NUT spacetime,''
\href{https://arxiv.org/pdf/1908.04238.pdf}{[arXiv:1908.04238 [gr-qc]]}.
\bibitem{Bordo:2019tyh}
A.~B.~Bordo, F.~Gray, R.~A.~Hennigar and D.~Kubiz\v{n}\'ak,
``Misner Gravitational Charges and Variable String Strengths,''
Class. Quant. Grav. \textbf{36}, no.19, 194001 (2019)
doi:10.1088/1361-6382/ab3d4d
[arXiv:1905.03785 [hep-th]].


\bibitem{Bordo:2019rhu}
A.~Ballon Bordo, F.~Gray, R.~A.~Hennigar and D.~Kubiz\v{n}\'ak,
``The First Law for Rotating NUTs,''
Phys. Lett. B \textbf{798}, 134972 (2019)
doi:10.1016/j.physletb.2019.134972
[arXiv:1905.06350 [hep-th]].

\bibitem{Zhang:2021pvx}
M.~Zhang and J.~Jiang,
``NUT charges and black hole shadows,''
Phys. Lett. B \textbf{816}, 136213 (2021)
doi:10.1016/j.physletb.2021.136213
[arXiv:2103.11416 [gr-qc]].

\bibitem{Lee:1990nz}
J.~Lee and R.~M.~Wald,
J. Math. Phys. \textbf{31}, 725-743 (1990)
doi:10.1063/1.528801
\bibitem{Iyer:1994ys}
V.~Iyer and R.~M.~Wald,
Phys. Rev. D \textbf{50}, 846-864 (1994)
doi:10.1103/PhysRevD.50.846
[arXiv:gr-qc/9403028 [gr-qc]].
\bibitem{Wald:1999wa}
R.~M.~Wald and A.~Zoupas,
Phys. Rev. D \textbf{61}, 084027 (2000)
doi:10.1103/PhysRevD.61.084027
[arXiv:gr-qc/9911095 [gr-qc]].

\bibitem{Barnich:2001jy}
G.~Barnich and F.~Brandt,
Nucl. Phys. B \textbf{633}, 3-82 (2002)
doi:10.1016/S0550-3213(02)00251-1
[arXiv:hep-th/0111246 [hep-th]].


\bibitem{Frodden:2019ylc}
E.~Frodden and D.~Hidalgo,
``Surface Charges Toolkit for Gravity,''
Int. J. Mod. Phys. D \textbf{29}, no.06, 2050040 (2020)
doi:10.1142/S0218271820500406
\href{https://arxiv.org/pdf/1911.07264.pdf}{[arXiv:1911.07264 [hep-th]]}.

\bibitem{metriccite}
Demianski M and Newman E T 1966 {\it Bull. Acad. Polon. Sci. Ser. Math. Astron. Phys.} {\bf 14} 653.

\bibitem{Manko:2006bf}
V.~S.~Manko, J.~Martin and E.~Ruiz,
Class. Quant. Grav. \textbf{23}, 4473-4484 (2006)
doi:10.1088/0264-9381/23/13/011
[arXiv:gr-qc/0603002 [gr-qc]].


\bibitem{podolskybook}
J.~B.~Griffiths and J.~Podolsky, {\it ``Exact Space-Times in Einstein’s General Relativity,''} doi:10.1017/CBO9780511635397.























\bibitem{Griffiths:2005qp}
J.~B.~Griffiths and J.~Podolsky,
Int. J. Mod. Phys. D \textbf{15}, 335-370 (2006)
doi:10.1142/S0218271806007742
[arXiv:gr-qc/0511091 [gr-qc]].





\end{thebibliography}
\end{document}